\def\ptitle{Energies and wave functions for a soft-core Coulomb potential}
\def\half{\frac{1}{2}}
\def\sgn{{\rm sgn}}
\def\dbox#1{\hbox{\vrule  
                        \vbox{\hrule \vskip #1
                             \hbox{\hskip #1
                                 \vbox{\hsize=#1}%
                              \hskip #1}%
                         \vskip #1 \hrule}%
                      \vrule}}
\def\qed{\hfill \dbox{0.05true in}}  
\documentclass{revtex4}
\usepackage{amsmath}
\usepackage{graphicx}
\begin{document}


\title{\ptitle}

\author{Richard L. Hall$^1$, Nasser Saad$^2$, K. D. Sen$^3$, and Hakan Ciftci$^4$\medskip }

\address{$^1$ Department of Mathematics and Statistics, Concordia University,
1455 de Maisonneuve Boulevard West, Montr\'eal, Qu\'ebec, Canada
H3G 1M8}
\address{$^2$ Department of Mathematics and Statistics,
University of Prince Edward Island, 550 University Avenue,
Charlottetown, PEI, Canada C1A 4P3.}
\address{$^3$ School of Chemistry, University  of Hyderabad 500046, India.}
\address{$^4$ Gazi Universitesi, Fen-Edebiyat Fak\"ultesi, Fizik
B\"ol\"um\"u, 06500 Teknikokullar, Ankara, Turkey.}

\email{rhall@mathstat.concordia.ca} \email{nsaad@upei.ca}
\email{sensc@uohyd.ernet.in} \email{hciftci@gazi.edu.tr}
\medskip
\begin{abstract}
For the family of model soft-core Coulomb potentials represented by
$V(r) = -\frac{Z}{(r^q+\beta^q)^{\frac{1}{q}}}$, with the
parameters
 $Z>0,~\beta>0,~q \ge 1$, it is shown analytically that the
potentials and eigenvalues, $E_{\nu\ell}$, are monotonic in each
parameter. The potential envelope method is applied to obtain
approximate analytic estimates in terms of the known exact spectra
for pure power potentials. For the case $q =1$, the Asymptotic
Iteration Method is used to find exact analytic results for the
eigenvalues $E_{\nu\ell}$  and corresponding wave functions,
expressed in terms of $Z$ and $\beta$. A proof is presented
establishing the general concavity of the scaled electron density
near the nucleus resulting from the truncated potentials for all
$q$. Based on an analysis of extensive numerical calculations, it
is conjectured that the crossing between the pair of states
$[(\nu,\ell),(\nu',\ell')]$, is given by the condition $\nu'\geq
(\nu+1)$ and $\ell' \geq (\ell+3).$ The significance of these results
 for the interaction of an intense laser field with an atom is pointed
out. Differences in the observed level-crossing effects between
the soft-core potentials and the hydrogen atom confined inside an
impenetrable sphere are discussed.

 \end{abstract}
 \maketitle

\noindent {\bf PACS:} 31.15.-p, 31.10.+z; 36.10.Ee; 36.20.Kd;
03.65.Ge.

\vskip0.2in
\noindent keywords: Soft-core Coulomb potential, asymptotic iteration
method, level crossing, Cusp condition, Intense laser atom
interaction,confined hydrogen atom,eigenvalues, eigenfunctions.

\maketitle
\section{Introduction}
The Schr\"odinger's time-independent equation  $H\Psi = E\Psi$,
 where the Hamiltonian $H$ is given by (in
atomic units $m = \hbar = e = 1$)
\begin{equation}\label{Eq:sro}
H = -\half\Delta + V_q(r),\quad V_q(r) =
-\frac{Z}{(r^q+\beta^q)^{\frac{1}{q}}}.
\end{equation}
introduces the family of soft-core (truncated) Coulomb potentials,
$V_q$, useful as model potentials in atomic physics. The bound
states are obtained in terms of three potential parameters: the
coupling $Z>0,$ the cutoff parameter $\beta >0,$ and the power
parameter $q \ge 1.$ The specific potentials corresponding to $q =
1$ and $q=2$ have been analyzed earlier
\cite{MP,P1,SVD,DMVD,SR,FF0,CM,OM,MO}. The potential $V_1$
represents the potential due to a smeared charge and may be useful
in describing mesonic atoms. The potential $V_2$ is similar to the
shape of the potential due to a finite nucleus and experienced by
the muon in a muonic atom. Extensive applications of the soft-core
Coulomb potential, $V_2$, have been made through model
calculations to describe the interaction of intense laser fields
with atoms \cite{LM,JHE1,JHE2,JHE3,JHE4,PLK,CWC,CK}. The parameter
$\beta$ can be related to the strength of the laser field, with
the range $\beta=20-40$ covering the experimental laser field
strengths \cite{LM}. Mehta and Patil \cite{MP} have presented
analytical solutions for the $s$-state eigenvalues corresponding
to the $V_1$ potential. Patil \cite{P1} has also discussed the
analyticity of the scattering phase shifts for two particles
interacting through the potentials
 $V_q$ with $q=1$ and $q=2$. Singh et al \cite{SVD} have reported
 a large number of eigenvalues for the states $1s$ to $4f$ corresponding to
 $V_1$ and $V_2$ for a fixed value of $Z$; these values were obtained by
 the numerical solution of Eq.(1) for $Z = 1;$ scaling laws,
 to be discussed here in section~2, extend their application to other values of $Z.$
 These authors noted that
 the energy-level ordering satisfied the condition $E_{\nu\ell} >
 E_{\nu\ell^{'}}$, where ${\ell} < {\ell^{'}}$. In these formulas, $\nu$ is the 'principal
quantum number', defined generally (also for non-Coulombic
potentials) as $\nu=n+\ell,$ where $n$ is the number of radial
nodes plus one.
 For each $\ell$
 value, the calculated energies were found to be well represented
 by a Ritz type formula. Exact bound-state solutions of $V_1$
 have been considered earlier \cite{DMVD,SR,FF0,CM}, wherein only a
 limited number of states with a specific choice of ${\ell}=0\dots 3$ have been treated.
 To our knowledge, no such study for $q \geq 2 $ has been
 reported so far. Further, the interesting possibility of realizing
 the condition $E_{\nu\ell}\equiv E_{\nu^{'}\ell^{'}}$ at a common value of $\beta$
 when $\nu^{'} > \nu $, and ${\ell}\neq{\ell^{'}}$ has not been considered as yet, for any $V_q$ .
In view of these observations based on the review of the previous
work reported on $V_q$, we have carried out a general analysis of
the characteristic features of the energies and wave functions of
the complete family of soft-core Coulomb potentials defined by $V_q$ as
a function of \emph{all} its parameters. Next, the potential
envelope method \cite{env1} is employed to express approximate
estimates of $E_{\nu\ell}$ showing an interesting geometric
property for all $V_q$. Our choice of the Asymptotic Iteration
Method (AIM) \cite{cs,BHS,ff,hs,ns,ba,bam,br,bb,sh} enables us to
present {\it general} new analytical results on exact bound-state
solutions corresponding to $V_1$. It is then shown analytically
that the electron density near the nucleus generated from all
$V_q$ is \emph{always} concave. Finally, the first numerical
results, on the crossing of energy levels in the energy spectrum
corresponding to $V_q,$ with $q=1$ and $q=2$ are reported. The
paper is organized as follows: scaling and montonicity laws are
established in sections~2 and 3; analytical spectral bounds are
found by means of envelope methods in section~4; in
sections~5~to~7 the Asymptotic Iteration Method is summarized and
used for the case $q  = 1$ to find exact analytical expressions
 for both eigenvalues and wave functions; in section~8 the concavity of the scaled electron density is established
 for all $q \ge 1$;
 in section~9 we discuss the characteristics of the crossings of the energy levels for
 soft-core Coulomb potentials, and also some comparisons of these results
with those for atoms confined inside an impenetrable sphere.
\section{Scaling}
The radial equation corresponding to (\ref{Eq:sro}) may be written
\begin{equation}\label{Eq:radial}
H\psi(r) = -\half\psi''(r) + \left(\frac{\ell(\ell+1)}{2r^2} -
\frac{Z}{(r^q+\beta^q)^{\frac{1}{q}}}\right)\psi(r) = E\psi(r),
\end{equation}
where $\psi(0) = 0.$ We shall take $Z$ to be a  positive real
parameter and express the general parametric dependence of the
eigenvalues in the form $E = E_{\nu\ell}(Z,\beta,q),$ in which, as
we have noted above, for a given $\ell,$ the `principal quantum
number' $\nu$ is defined by $\nu = n+\ell,$ where $n$ is the
number of radial nodes plus one.
 If we make the change of variables $r\rightarrow \sigma
r$ in (\ref{Eq:radial}), where $\sigma >0$ is constant, multiply
through by $\sigma^2$, and compare eigenvalues, we  immediately
arrive at the general scaling law for this class of potentials,
namely:
\begin{equation}\label{Eq:scaling}
E(Z,\beta,q) = \frac{1}{\sigma^2}E(\sigma Z, \beta/\sigma,q).
\end{equation}
The two special cases $\sigma = 1/Z$ and $\sigma = \beta$ then
yield, respectively, the special scaling laws
\begin{equation}\label{Eq:scalings}
E(Z,\beta,q) = Z^2E(1,Z\beta,q) = \frac{1}{\beta^2}E(Z\beta,1,q).
\end{equation}
The parameter $q$ is not involved because the denominator of the
potential always scales like length, for every $q > 0.$
\section{Monotonicities}
The eigenvalues $E_{\nu\ell} = E(Z,\beta,q)$ are montone in each
of the three potential parameters.  In fact we shall now show:
\begin{equation}\label{Eq:monotone}
\frac{\partial E}{\partial Z} < 0,\quad \frac{\partial E}{\partial
\beta} > 0,\quad {\rm and}\quad \frac{\partial E}{\partial q} < 0.
\end{equation}
The Schr\"odinger operator $H$ is bounded below. This may be shown
by an application of the operator inequality \cite{RS2,GS}
$-\Delta > 1/(4r^2)$ which yields the general spectral bound $E >
\min_{r > 0}[1/(8r^2) + V(r)].$. Explicit upper and lower bounds
for all the eigenvalues may be expressed in this form with the aid
of the `potential envelope method'
\cite{env1,env2,env3,env4,env5,env6,env7}; this will be discussed
in the next section.  Thus the discrete spectrum of $H$ may be
characterized variationally, and from this it follows that
monotonicities in the potential's dependence  on the parameters
induces the same monotonicities in the eigenvalues.  We therefore
prove (\ref{Eq:monotone}) by establishing, in turn, the
corresponding monotonicities in the potential.  First, by
inspection, we see immediately that $\partial V/\partial Z < 0.$
In what follows it is convenient to write $V(r) = -Z/F(r),$ and to
note that, if $s$ is a potential parameter, then
\[
\frac{\partial V}{\partial s} = \frac{Z}{F^2(r)}\frac{\partial
F}{\partial s}.
\]
Thus $\partial V/\partial s$, $\partial F/\partial s$, and
$\partial G/\partial s$ have the same sign, where $G(r) =
\ln(F(r))$ is given by
\[
G(r) = \frac{1}{q}\ln(r^q + \beta^q).
\]
We see that $\partial G/\partial \beta = \beta^{q-1}/(r^q +
\beta^q) > 0$; hence we have $\partial E/\partial \beta > 0.$
Finally, we consider $\partial G/\partial q.$  We first suppose
that $r < \beta$ and we define $x = r/\beta < 1.$ In terms of this
new variable we find
\[
\frac{\partial G}{\partial q} = -\frac{1}{q^2}\ln(1 + x^q) +
\frac{1}{q}\left(\frac{x^q\ln(x)}{1+x^q}\right).
\]
Since $0< x< 1,$ it follow that $\partial G/\partial q < 0$ for $r
< \beta.$  An exactly analogous argument with $y = \beta/r$ for $r
\ge\beta$ shows that $\partial G/\partial q < 0$ when $r \ge
\beta.$ Thus we conclude that $\partial E/\partial q < 0.$  We
note parenthetically that this result is in contrast to a
generalized mean $M(q)$ of $r$ and $\beta$ given, for example, by
\[
M(q) = \left(\frac{r^q + \beta^q}{2}\right)^{\frac{1}{q}}.
\]
Such a mean is known \cite{HLP} to be montone {\it increasing} in
$q$: the `$2$' in the denominator of the expression for $M(q)$
makes the difference.
\medskip

\begin{figure}[htbp]\centering\includegraphics[width=12cm]{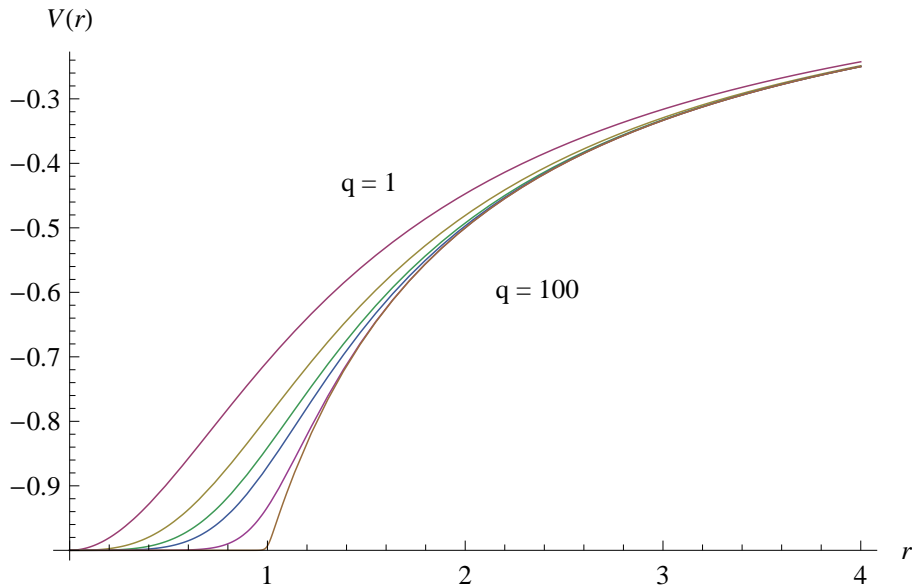}
\caption{The family of potentials $V(r) = -1/(r^q + 1)^q$ for
$1\leq q < \infty$ (in $a.u.$). The graphs are non-intersecting
for $r >0$, and are ordered, decreasing as $q$
increases.}\label{potentials}\end{figure}

The analysis of the monotonicity of the potential with respect to
$q$ may be  used to find the limiting potential as $q\rightarrow
\infty.$ We have for $r < \beta$ and $x = r/\beta < 1,$
\[
\lim\limits_{q\rightarrow\infty}G(r) =
\lim\limits_{q\rightarrow\infty}\left[\ln(\beta) +
\frac{\ln(1+x^q)}{q}\right] = \ln(\beta).
\]
Similarly, for $r \ge \beta$ and $y = \beta/r \le 1,$ we find
 \[
\lim\limits_{q\rightarrow\infty}G(r) =
\lim\limits_{q\rightarrow\infty}\left[\ln(r) +
\frac{\ln(1+y^q)}{q}\right] = \ln(r).
\]
We conclude
\begin{equation}\label{Eq:limV}
\lim_{q\rightarrow\infty} V(r) = V_{\infty}(r) = \left\{
\begin{array}{l l}
-\frac{Z}{\beta},  &\text{if $r < \beta$;}\\
\\
-\frac{Z}{r},      &\text{if $r \ge \beta$.}
\end{array}
\right.
\end{equation}
Thus, for given $Z$ and $\beta,$ the family of potentials
$\{V(r)\}_{q = 1}^{\infty}$ has an ordered set of graphs, of which
$V_{\infty}(r)$ is the lowest: they intersect only at $r = 0.$ For
$Z = \beta = 1,$ the family of potentials is illustrated in
Fig.~(\ref{potentials}).
\section{Energy bounds by the `potential envelope method'}
As we mentioned above, the operator inequality $-\Delta >
1/(4r^2)$ proved in Refs. \cite{RS2,GS} immediately  yields a
lower energy bound that is expressed by a classical formula,
namely
\[
E > \min_r\,\left[\frac{1}{8r^2} + V(r)\right].
\]
The potential envelope method allows us to construct tighter and
more specific energy formulas of this type.  The method explores
the idea that, if a given potential $V(r)$ can be written as a
smooth transformation $V(r) = g(h(r))$ of a potential $h(r)$, for
which the spectral problem is solved, then the Schr\"odinger
spectrum generated by $V(r)$ may be expressed in terms of the
spectrum associated with the basis potential $h(r).$  The method
was introduced \cite{env1} in 1980  as a technique for the
many-body problem and has subsequently been developed and  used
also for single-particle problems \cite{env2,env3,env4}, including
the laser-dressed potential \cite{env5}, which we call $V_2(r)$ in
the present article, and for relativistic problems
\cite{env6,env7}. If the transformation function $g(h)$ is convex,
the method yields lower energy bounds; conversely, when $g(h)$ is
concave, the results are upper bounds.  If the basis $h(r)$ of the
transformation is a pure power $h(r) = \sgn(p)\,r^p$, then the
resulting energy formula has the form
\begin{equation}\label{env}
E_{\nu\ell} =  \min_{r >0}\left[\frac{P_{\nu\ell}^2(p)}{2r^2} -
\frac{Z}{(r^q+\beta^q)^\frac{1}{q}}\right],
\end{equation}
where $P_{\nu\ell}(p)$ is a constant determined by $h(r).$ For the
class of soft-core Coulomb potentials $V_q(r)$ we study here, we are
able to obtain lower bounds if $h(r) = -1/r,$ and  upper bounds if
$h(r) = r^2.$ With the Hydrogenic basis $p = -1$, $g(h)$ is convex
for all $q \ge 1$, and $P_{\nu\ell}(-1) = \nu.$ For the oscillator
basis $p = 2,$ we have $P_{\nu\ell}(2) = 2\nu -(\ell+\half)$ and
the transformation $g(h)$ is concave for $q <= 2;$ for larger $q$,
in this case, the formula (\ref{env}) yields an upper bound
provided the critical $r  = \hat{r}$ found is not too small,
namely for $q = \{3,4,5,6\},$ respectively, we require
$\hat{r}/\beta \ge \{0.958, 1.233,1,356,1.417\}.$ This condition
arises because, even though $g(h)$ in this case ($h = r^2$, $q >
2$) is  not concave everywhere, tangential potentials of the form
$a + br^2$, which touch in the concave region, can still provide
upper bounds provided they do not cross $V_q(r) = g(r^2).$ These
energy bounds are in effect exploratory tools. The bounds obey the
same scaling and monotonicity properties as the exact eigenvalues.
It is perhaps geometrically interesting that the entire set of
upper and lower approximate energy curves, expressing $E$ as a
function of the coupling $Z$, are magnifications of a single
curve, with magnification factor $P^2$: if we write $V_q(r) = Z
f(r),$
 then the parametric equations for the energy-bound curves are given by

\begin{eqnarray}\label{para}
\left\{ \begin{array}{ll}
  Z = P^2\left[\frac{1}{r^3f'(r)}\right]\\
\\
 E = P^2\left[\frac{1}{2r^2} + \frac{f(r)}{r^3f'(r)}\right].
       \end{array} \right.
\end{eqnarray}

\section{The asymptotic iteration method}
\noindent The asymptotic iteration method (AIM) was original
introduced \cite{cs} to investigate the solutions of differential
equations of the form
\begin{equation}\label{aimsf}
y''=\lambda_0(r) y'+s_0(r) y,\quad\quad ({}^\prime={d\over dr})
\end{equation}
where $\lambda_0(r)$ and $s_0(r)$ are $C^{\infty}-$differentiable
functions. A key feature of this method is to note the invariant
structure of the right-hand side of (\ref{aimsf}) under further
differentiation. Indeed, if we differentiate (\ref{aimsf}) with
respect to $r$, we obtain
\begin{equation}\label{aimfd}
y^{\prime\prime\prime}=\lambda_1 y^\prime+s_1 y
\end{equation}
where $\lambda_1= \lambda_0^\prime+s_0+\lambda_0^2$ and
$s_1=s_0^\prime+s_0\lambda_0.$ If we find the second derivative of
equation (\ref{aimsf}), we obtain
\begin{equation}\label{aimsd}
y^{(4)}=\lambda_2 y^\prime+s_2 y
\end{equation}
where $\lambda_2= \lambda_1^\prime+s_1+\lambda_0\lambda_1$ and
$s_2=s_1^\prime+s_0\lambda_1.$ Thus, for $(n+1)^{th}$ and
$(n+2)^{th}$ derivative of (\ref{aimsf}), $n=1,2,\dots$, we have
\begin{equation}\label{aimnf}
y^{(n+1)}=\lambda_{n-1}y^\prime+s_{n-1}y
\end{equation}
and
\begin{equation}\label{aimnpf}
y^{(n+2)}=\lambda_{n}y^\prime+s_{n}y
\end{equation}
respectively, where
\begin{equation}\label{aimseq}
\lambda_{n}=
\lambda_{n-1}^\prime+s_{n-1}+\lambda_0\lambda_{n-1}\hbox{ ~~and~~
} s_{n}=s_{n-1}^\prime+s_0\lambda_{n-1}.
\end{equation}
From (\ref{aimnf}) and (\ref{aimnpf}) we have
\begin{equation}\label{aimdelta}
\lambda_n y^{(n+1)}- \lambda_{n-1}y^{(n+2)} = \delta_ny {\rm
~~~where~~~}\delta_n=\lambda_n s_{n-1}-\lambda_{n-1}s_n.
\end{equation}
Clearly, from (\ref{aimdelta}) if $y$, the solution of
(\ref{aimsf}), is a polynomial of degree $n$, then $\delta_n\equiv
0$. Further, if $\delta_n=0$, then $\delta_{n'}=0$ for all $n'\geq
n$. In an earlier paper \cite{cs} we proved the principal theorem
of the Asymptotic Iteration Method (AIM), namely \vskip0.1in
\noindent{\bf Theorem 1:} \emph{Given $\lambda_0$ and $s_0$ in
$C^{\infty}(a,b),$ the differential equation (\ref{aimsf}) has the
general solution
\begin{equation}\label{aimgs}
y(r)= \exp\left(-\int\limits^{r}\alpha(t) dt\right) \left[C_2
+C_1\int\limits^{r}\exp\left(\int\limits^{t}(\lambda_0(\tau) +
2\alpha(\tau)) d\tau \right)dt\right]
\end{equation}
if for some $n>0$
\begin{equation}\label{aimcond}
{s_{n}\over \lambda_{n}}={s_{n-1}\over \lambda_{n-1}} \equiv
\alpha.
\end{equation}
}
\section{Exact energy eigenvalues}
\noindent Since the earlier work of Ciftci et al \cite{cs}, AIM
has been adopted to investigate a wide range of different problems
in relativistic and non-relativistic quantum mechanics. It should
be noted that, in the process of applying AIM, especially to
eigenvalue problems of Schr\"odinger-type, such as
(\ref{Eq:radial}), one has to overcome the following two problems
\cite{BHS}.

\subsection{Asymptotic Solution Problem} \noindent This problem deals with the conversion of the eigenvalue problem (the absence
of first derivative) to standard form (\ref{aimsf}) suitable for
the application of AIM. A general strategy to overcome this
problem is to use $\psi(r) \equiv \psi_a(r)f(r)$ where $\psi_a(r)$
is an asymptotic solution satisfying the boundary conditions of
the given eigenvalue equation and $f(r)$ is an unknown function to
be determine using AIM. For (\ref{Eq:radial}), we note that, as
$r$ approaches $\infty$, the asymptotic solution $\psi_\infty(r)$
of (\ref{Eq:radial}) satisfies the differential equation
 \begin{equation}\label{wfasyi}
\psi_\infty''(r) \approx -2 E\psi_\infty(r),
\end{equation}
which yields
 \begin{equation}\label{wfis}
\psi_\infty (r) \approx e^{-\sqrt{-2 E}r}.
\end{equation}
Meanwhile, as $r$ approaches $0$, the asymptotic solution
$\psi_0(r)$ of (\ref{Eq:radial}) satisfies the differential
equation
 \begin{equation}\label{wf0}
-\psi_0''(r) + \frac{\ell(\ell+1)}{r^2}\psi_0(r)\approx0,
\end{equation}
which assumes the solution
 \begin{equation}\label{wf0s}
\psi_0 (r) \approx r^{l+1}.
\end{equation}
Consequently, we may write the exact solution of (\ref{Eq:radial})
as
  \begin{equation}\label{wfgf}
\psi (r) = r^{l+1}e^{-\sqrt{-2 E}r}f(r).
\end{equation}
Substituting (\ref{wfgf}) into the Schr\"odinger equation
(\ref{Eq:radial}), we find that the radial function $f(r)$ must
satisfy the differential equation
  \begin{equation}\label{aimgfsl}
f''(r)=2\left(a-{l+1\over r}\right)f'(r)+\left({2(l+1)a\over
r}-{2Z\over (r^q+\beta^q)^{1\over q}}\right)f(r)
\end{equation}
where we denote
  \begin{equation}\label{energy}
a^2=-2E.
\end{equation}
AIM is then used to solve this second-order homogeneous
differential equation for $f(r)$.

\subsection{\bf Termination Condition Problem}\noindent This problem results when the eigenvalue problem (now in the standard form for an AIM application) fails to be exactly solvable. Indeed, if the eigenvalue problem has exact analytic solutions, the termination condition
(\ref{aimcond}), or equivalently,
  \begin{equation}\label{aimcondre}
\delta_n(r;E)=\lambda_n(r;E)s_{n-1}(r;E)-\lambda_{n-1}(r;E)s_n(r;E)\equiv
0
\end{equation}
produces, at each iteration, an expression that is independent of
$r$. For example, if $\beta = 0$, Eq.(\ref{aimgfsl}) is exactly
solvable and the termination condition (\ref{aimcondre}) yields
\begin{equation}\label{aimcoul}
\delta_n(E)=\prod\limits_{k=0}^{n+1}(-Z+(k+l)a),\quad n=1,2,\dots
\end{equation}
Thus the AIM condition $\delta_n(E) = 0$ leads to
\begin{equation}\label{coulexact}
E_n=-\half{Z^2\over (n+l)^2},\quad n=1,2,\dots,
\end{equation}
as expected for the exact solutions of Coulomb potential. For
$\beta\neq 0$, Eq.(\ref{aimgfsl}) is not exactly solvable in
general, however, for certain values of the potential parameters,
we can obtain some analytic solutions. For example, in the case of
$q=1$, we may use AIM with
\begin{equation}\label{lambda0}
\lambda_0=2\left(a-{l+1\over r}\right),\quad
s_0=\left({2(l+1)a\over r}-{2Z\over r+\beta}\right),
\end{equation}
to obtain a class of exact solutions given by
\begin{equation}\label{tcoulenergy}
a={Z\over l+n+1},
\end{equation}
where $n=1,2,3,\dots$ is the iteration number used by AIM, along
with the conditions on the potential parameter $\beta$ reported in
Table 1.

\begin{table}
\caption{\label{tab:table1} Exact solutions of the radial
Schr\"odinger equation (\ref{Eq:radial}) for $q=1$.}
\begin{ruledtabular}
\begin{tabular}{l|l}
 $a$& Conditions on $\beta$\\
\hline
$\frac{Z}{l+2}$& $-Z\beta+l+2=0$ \\
\\
$\frac{Z}{l+3}$& $Z^2(l+2)\beta^2-3Z(l+2)(l+3)\beta+(2l+3)(l+3)^2=0$ \\
\\
$\frac{Z}{l+4}$& $-Z^3(l+3)(l+2)\beta^3+6Z^2(l+4)(l+3)(l+2)\beta^2-Z(11l^2+50l+54)(l+4)^2\beta$\\&$+3(l+2)(2l+3)(l+4)^3=0$ \\
\\
$\frac{Z}{l+5}$& $Z^4(l+4)(l+3)(l+2)\beta^4-10Z^3(l+5)(l+4)(l+3)(l+2)\beta^3$\\
&$+Z^2(35l^3+300l^2+823l+720)(l+5)^2\beta^2-Z(50l^3+381l^2+925l+720)(l+5)^3\beta$\\
&$+6(2l+5)(2l+3)(l+2)(l+5)^4=0$\\
\\
$\frac{Z}{l+6}$& $-Z^5(l+5)(l+4)(l+3)(l+2)\beta^5+15Z^4(l+6)(l+5)(l+4)(l+3)(l+2)\beta^4$\\
&$-Z^3(85l^4+1155l^3+5678l^2+11928l+9000)(l+6)^2\beta^3$\\
&$+3Z^2(75l^4+952l^3+4359l^2+8522l+6000)(l+6)^3\beta^2$\\
&$-Z(274l^4+3073l^3+12411l^2+21492l+13500)(l+6)^4\beta$\\
&$+30(2l+5)(2l+3)(l+3)(l+2)(l+6)^5=0$\\
\\
$\frac{Z}{l+7}$& $Z^6(l+6)(l+5)(l+4)(l+3)(l+2)\beta^6-21Z^5(l+7)(l+6)(l+5)(l+4)(l+3)(l+2)\beta^5$\\
&$+Z^4(175l^5+3430l^4+26033l^3+95354l^2+167976l+113400)(l+7)^2\beta^4$\\
&$-3Z^3(245l^5+4592l^4+33271l^3+116224l^2+195300l+126000)(l+7)^3\beta^3$\\
&$+Z^2(1624l^5+28182l^4+188607l^3+608332l^2+945783l+567000)(l+7)^4\beta^2$\\
&$-9Z(196l^5+3004l^4+17753l^3+50746l^2+70301l+37800)(l+7)^5\beta$\\
&$+90(2l+7)(2l+5)(2l+3)(l+3)(l+2)(l+7)^6=0$\\
\\
$\frac{Z}{l+8}$& $-Z^7(l+7)(l+6)(l+5)(l+4)(l+3)(l+2)\beta^7+28Z^6(l+8)(l+7)(l+6)(l+5)(l+4)(l+3)(l+2)\beta^6$\\
&$-2Z^5(161l^6+4284l^5+46109l^4+256284l^3+773558l^2+1198080l+740880)(l+8)^2\beta^5$\\
&$+2Z^4(980l^6+25263l^5+263144l^4+1414449l^3+4127804l^2+6184116l+3704400)(l+8)^3\beta^4$\\
&$-Z^3(6769l^6+165501l^5+1632238l^4+8299620l^3+22916602l^2+32533488l+18522000)(l+8)^4\beta^3$\\
&$+2Z^2(6566l^6+148023l^5+1343681l^4+6287868l^3+16004408l^2+21011364l+11113200)(l+8)^5\beta^2$\\
&$-9Z(1452l^6+28968l^5+232875l^4+968195l^3+2199048l^2+2589112l+1234800)(l+8)^6\beta$\\
&$+630(2l+7)(2l+5)(2l+3)(l+4)(l+3)(l+2)(l+8)^7=0$\\
\\
$\frac{Z}{l+9}$& $Z^8(l+8)(l+7)(l+6)(l+5)(l+4)(l+3)(l+2)\beta^8-36Z^7(l+9)(l+8)(l+7)(l+6)(l+5)(l+4)(l+3)(l+2)\beta^7$\\
&$+6Z^6(91l^7+3150l^6+45472l^5+354060l^4+1601869l^3+4198770l^2+5883828l+3386880)(l+9)^2\beta^6$\\
&$-18Z^5(252l^7+8519l^6+120015l^5+911495l^4+4021353l^3+10279346l^2+14054940l+7902720)(l+9)^3\beta^5$\\
&$+3Z^4(7483l^7+243355l^6+3294188l^5+24020450l^4+101717325l^3+249667695l^2+328201704l+177811200)(l+9)^4\beta^4$\\
&$-18Z^3(3738l^7+114755l^6+1464128l^5+10054742l^4+40105738l^3+92835203l^2+115350696l+59270400)(l+9)^5\beta^3$\\
&$+Z^2(118124l^7+3338080l^6+39154679l^5+247223313l^4+907897077l^3+1939695507l^2+2232161820l+1066867200)(l+9)^6\beta^2$\\
&$-18Z(6088l^7+152716l^6+1592078l^5+8960617l^4+29441156l^3+56502567l^2+58655178l+25401600)(l+9)^7\beta$\\
&$+2520(2l+9)(2l+7)(2l+5)(2l+3)(l+4)(l+3)(l+2)(l+9)^8=0$\\
\end{tabular}
\end{ruledtabular}
\end{table}

\section{Wave functions}
\noindent In this section we examine the properties of the wave
functions associated with the exact eigenvalues obtained in the
previous section. We note first, for $a={Z\over l+2}$ and
$\beta={l+2\over Z}$, the nodeless wave function of
(\ref{Eq:radial}) is given by
\begin{equation}\label{wfcase1}
\psi(r)=r^{l+1}~e^{-ar}~(1+{r\over \beta}).
\end{equation}
 Further, for $a={Z\over l+3}$,
we have the following result. \vskip0.1true in \noindent{\bf
Theorem:} For the radial Schr\"odinger equation (\ref{Eq:radial}),
if
\begin{equation}\label{betav1}
\beta={(l+3)[(6+3l)+\sqrt{(2+l)(6+l)}]\over 2(l+2)Z}
\end{equation}
then the exact wave function
\begin{align}\label{wfexact}
\psi(r)&=r^{l+1}~e^{-ar}\left(1+\left({\beta(l+3)(2l+3)Z+\sqrt{\beta(l+3)^2(2l+3)Z(4(l+3)^2-b(2l+5)Z)}\over 2\beta(l+3)^2(2l+3)}\right)r\right)\notag\\
&\times
\left(1+\left(a-{\beta(l+3)(2l+3)Z+\sqrt{\beta(l+3)^2(2l+3)Z(4(l+3)^2-\beta(2l+5)Z)}\over
2\beta(l+3)^2(2l+3)} \right)r\right),
\end{align}
is nodeless, while, if
\begin{equation}\label{betav2}
\beta={(l+3)[(6+3l)-\sqrt{(2+l)(6+l)}]\over 2(l+2)Z},
\end{equation}
then the exact wave function of (\ref{wfexact}) has exactly one
node located at
\begin{equation}\label{rlocat}
r\equiv-\half{(l+3)(2l+3)\over
(l+2)(l-\sqrt{(l+2)(l+6)})}\left[3(l+2)-\sqrt{(l+2)(l+6)}+\sqrt{2(l+2)(l+4+\sqrt{(l+2)(l+6)})}\right].
\end{equation}

\noindent{Proof:} Substitute $f(r)=(1+sr)(1+tr)$ in
(\ref{aimgfsl}) and equate the coefficients of $r$'s, we find that
$a={Z\over l+3}$ and $s=a-t$ where $t$ satisfies
$$t^2-at+ {(l+2)Z^2\over (3+2l)(l+3)^2}-{Z\over (3+2l)\beta}=0.$$
Further, $f(r)$ now reads
\begin{align*}
f(r)&=1+ar+(at-t^2)r^2=1+ar+\left[{(l+2)Z^2\over
(3+2l)(l+3)^2}-{Z\over (3+2l)\beta}\right]r^2.
\end{align*}
From Descartes' rule of signs, $f(r)$ has no real roots if
$\left[{(l+2)Z^2\over (3+2l)(l+3)^2}-{Z\over
(3+2l)\beta}\right]>0$ which holds for $\beta$ given by
(\ref{betav1}) and has only one positive root if
$\left[{(l+2)Z^2\over (3+2l)(l+3)^2}-{Z\over
(3+2l)\beta}\right]<0$ which holds true for $\beta$ given by
(\ref{betav2}).\qed

\noindent In Figures 2 and 3, we plot the un-normalized wave
functions for $a={Z\over l+3}$ and $\beta$ given by (\ref{betav1})
and (\ref{betav2}), respectively.
\begin{figure}[htbp]
\centering\includegraphics[width=3cm,height=3cm]{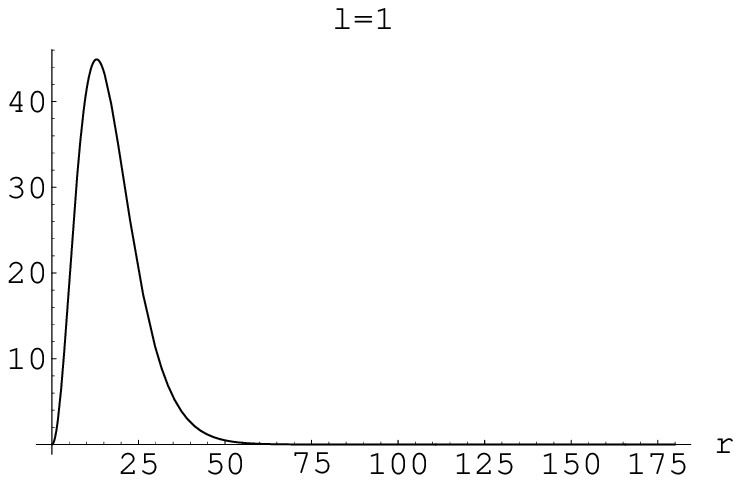}
\includegraphics[width=3cm,height=3cm]{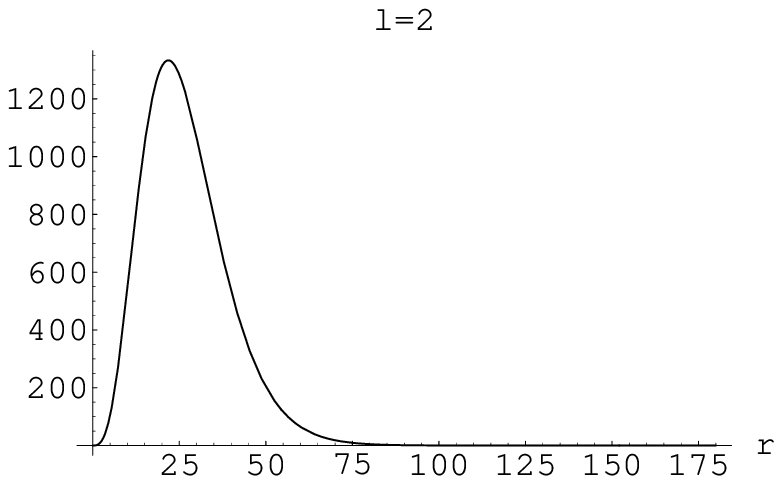}
\includegraphics[width=3cm,height=3cm]{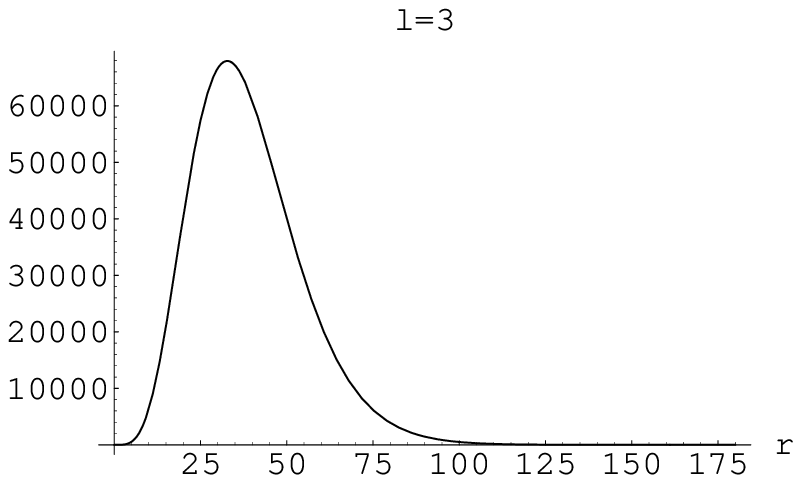}
\includegraphics[width=3cm,height=3cm]{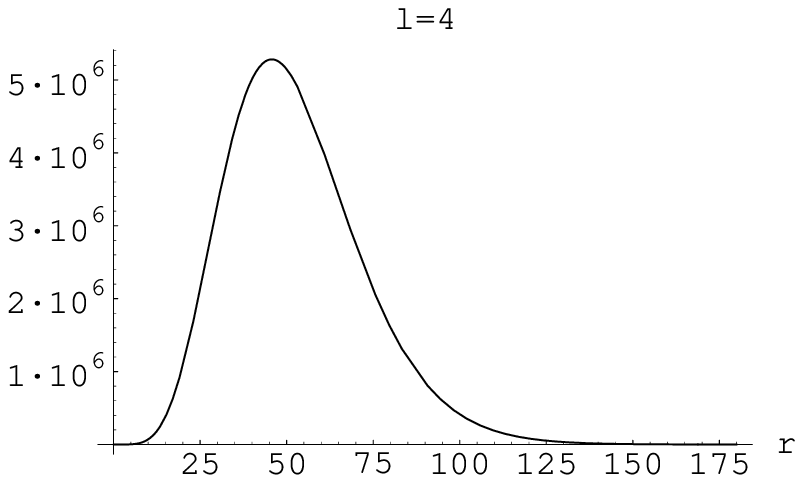}
\includegraphics[width=3cm,height=3cm]{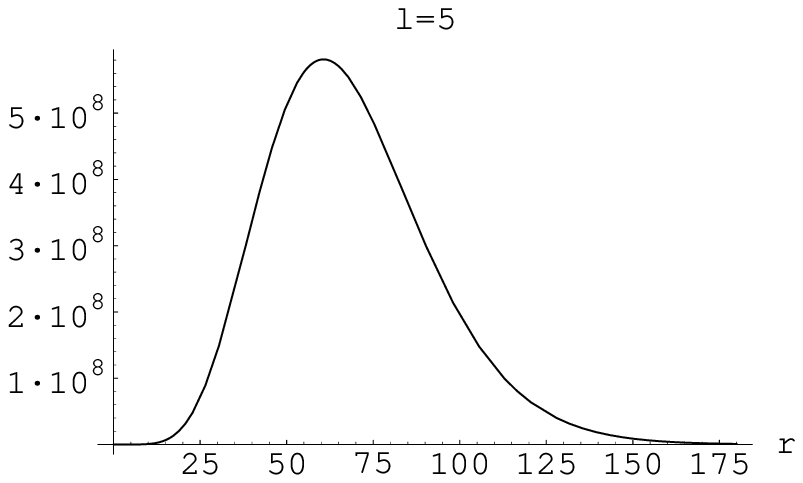}
\caption{Un-normalized wave functions (\ref{wfexact}) for the
Schr\"odinger equation,  Eq.(\ref{Eq:radial}), with soft-core Coulomb
potential, $q=1$, where $a={1\over l+3}$ and $\beta$ given by
(\ref{betav1}) for different values of $l$~(in
$a.u.$).}\label{wf1}\end{figure}
\begin{figure}[htbp]
\centering\includegraphics[width=3cm,height=3cm]{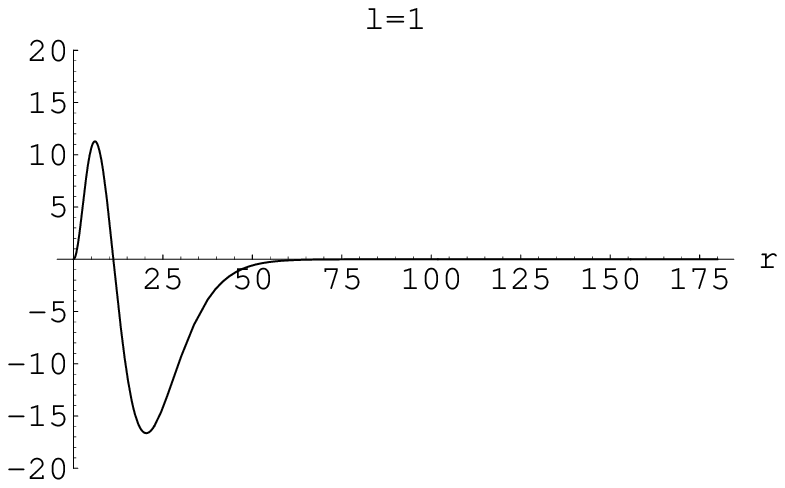}
\includegraphics[width=3cm,height=3cm]{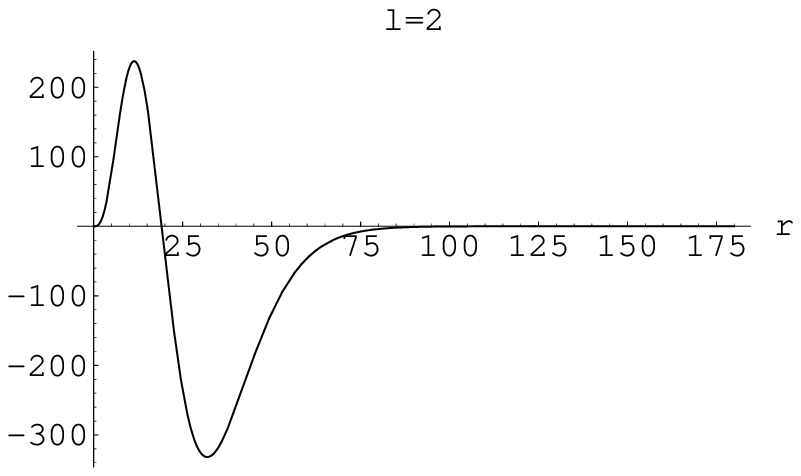}
\includegraphics[width=3cm,height=3cm]{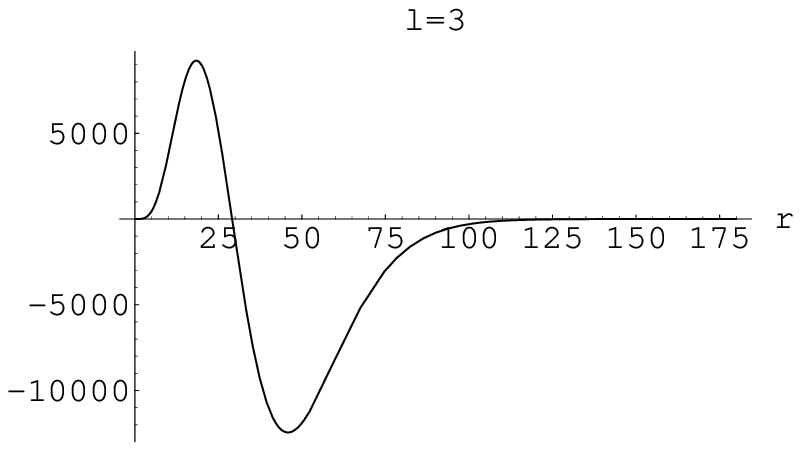}
\includegraphics[width=3cm,height=3cm]{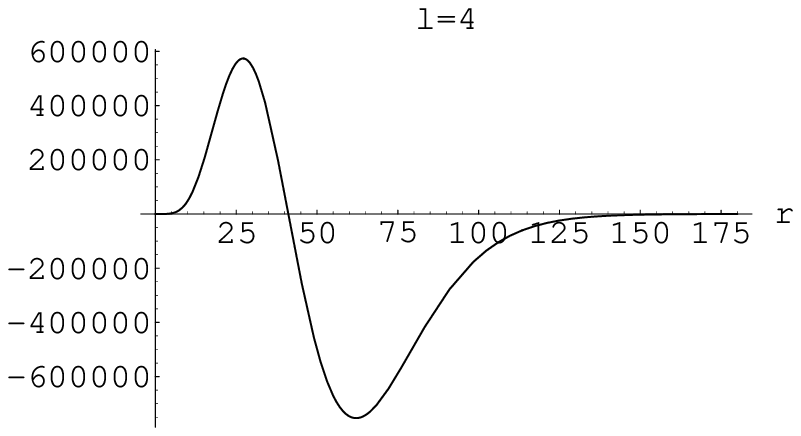}
\includegraphics[width=3cm,height=3cm]{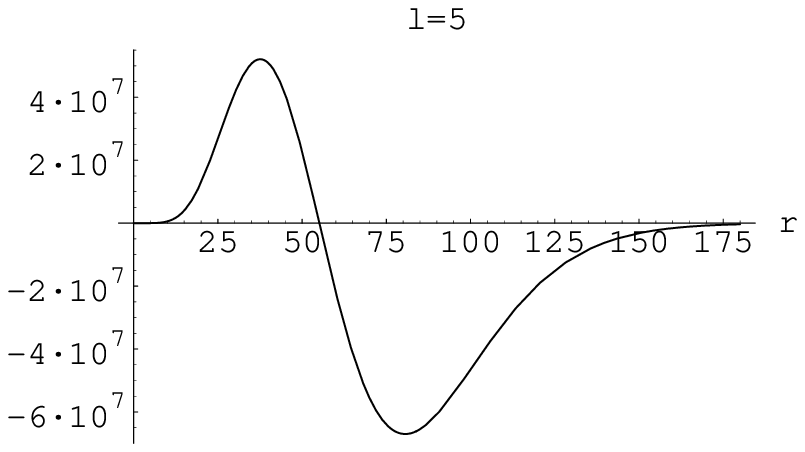}
\caption{Un-normalized wave functions (\ref{wfexact}) for the
Schr\"odinger equation,  Eq.(\ref{Eq:radial}), with soft-core Coulomb
potential, $q=1$, where $a={1\over l+3}$ and $\beta$ given by
(\ref{betav2}) for different values of $l$~(in $a.u.$).}
\label{wf2}\end{figure} Further results can be obtained similarly.
Indeed, for $a={Z\over (l+4)}$, it is straightforward to show that
the solution of (\ref{aimgfsl}) takes the form
$f(r)=(1+(a-k-t)r)(1+kr)(1+tr)$ with exactly no root, one root and
two roots, corresponding to the positive zeros of
$$-Z^3(l+3)(l+2)\beta^3+6Z^2(l+4)(l+3)(l+2)\beta^2-Z(11l^2+50l+54)(l+4)^2\beta+3(l+2)(2l+3)(l+4)^3=0.$$
\section{Concavity of the scaled electron density near the nucleus}
\noindent Using the radial wave function for the $\nu l$ state
given by
\begin{eqnarray}
R_{\nu l}(r)= \frac{\psi_{\nu l}(r)}{r^{2l+1}}~,
\end{eqnarray}
with the spherically averaged density normalized according to
\begin{eqnarray}
4 \pi \int {R_{\nu l}}^2 (r)r^2 dr = 4 \pi \int{\overline{\varrho
}(r)}r^2 dr = 1~,
\end{eqnarray}
we define the \emph{scaled} density as
\begin{eqnarray}
\eta_l (r) =  \frac {\overline{\varrho} (r)}{r^{2l}}
\end{eqnarray}
We note here that the Kato-Steiner cusp condition
\cite{KS1,KS2,PB,NS1} in terms of the density is given by
\begin{eqnarray}
\eta_l^{\prime} (0)  =  -  \frac {2Z}{l+1} \eta_l (0)
\end{eqnarray}

Now, for a spherical potential of the form $-\frac {A}{r}+B+f(r)$,
where $f(r)\rightarrow 0 ~as~ r\rightarrow 0$ it has been shown
\cite{SM1,SPM} that
\begin{equation}\label{etapp}
\eta_l^{\prime\prime} (0) =  \frac {2}{2l+3} \left[ \frac {A^2}{(
l+1 )^2} (4l+5) +2(B-E_{nl} )\right] \eta_l (0)
\end{equation}
Writing the soft-core Coulomb potential as
\[
V_q(r)= -\frac {Z}{\beta} + \left(\frac{Z}{\beta}
-\frac{Z}{(r^q+\beta^q)^{\frac{1}{q}}}\right) = B + f(r),
\]
we see that $A = 0$, $B = -Z/\beta,$ and $f(0) = 0.$ Thus we have
from (\ref{etapp})
\begin{eqnarray}
\eta_l^{\prime\prime} (0) = - \frac {4}{2l+3}\left[E_{\nu l} +
\frac{Z}{\beta}\right] \eta_l (0)
\end{eqnarray}
The sign of $\eta_l^{\prime\prime} (0)$ is then completely
determined by the quantity $E_{\nu l} + \frac{Z}{\beta}$. Writing
\begin{eqnarray}
E_{\nu l}= (\psi_{\nu l},H\psi_{\nu l}) =~\left(\psi_{\nu
l},\left[-\frac{1}{2}\Delta + V_q(r)\right]\psi_{\nu l}\right),
\end{eqnarray}
we find that
\begin{eqnarray}
E_{\nu l}\geq \left(\psi_{\nu l}, V_q(r)\psi_{\nu l}\right)\geq
\min_r\,\left[V_q(r)\right]= -\frac{Z}{\beta}
\end{eqnarray}
Therefore, for all $V_q(r), ~\eta_l^{\prime\prime} (0)\leq 0$ i.e.
the electron density near the nucleus remains \emph{concave}.
Specifically for $V_1(r)$, the exact nodeless wave functions at
$\beta= \frac{(l+2)}{Z}$ correspond to $E = -\frac{Z^2}{2(l+2)^2}$
which leads to the simple result that $$
\frac{\eta_l^{\prime\prime} (0)}{\eta_l (0)}=-
\frac{2Z^2}{(l+2)^2}.$$ The corresponding ratios for the results
obtained in Table I, given by AIM, can be similarly derived. Our
numerical analysis on the trends in the eigenvalues for the
potential $V_2(r)$, support the earlier result \cite{CWC} that the
nodeless wave functions with energy $E = -\frac{Z^2}{2(l+2)^2}$
are produced at $\beta=\sqrt \frac{2(l+2)^3}{Z^2}$. This leads to
the ratio $$ \frac{\eta_l^{\prime\prime} (0)}{\eta_l (0)}=-
\frac{2Z^2[\sqrt2(l+2) -1]}{(2l+3)(l+2)^2}.$$ Such ratios can be
obtained for the other $\beta$ values at which the exact energy
values are known \cite{CWC}. It has been argued \cite{EW,NH} that
the electron density in terms of the ratio $\frac{\eta_l^{\prime}
(0)}{\eta_l (0)}=  - \frac{2Z}{l+1} $ carries information on the
identity of the nuclei, i.e., the value of $Z$ and its location.
The corresponding external potential at the equilibrium density
thus provides a rationalization for the existence of the ground
state energy-density functional \cite{HK}. An interesting
situation arises in the case of the $V_q(r)$ class of potentials
when $\eta_l^{\prime} (0) $ vanishes. The information on the
identity of the nuclei is then contained in the ratio
$\frac{\eta_l^{\prime\prime} (0)}{\eta_l (0)}.$ We conclude this
section by noting that Eq.(38) can be used to obtain the condition
on $\frac{\eta_l^{\prime\prime} (0)}{\eta_l (0)}$ corresponding to
the critical potential at which $E_{\nu l}\equiv$ 0.
\section{Energy degeneracies for $\beta > 0$ and the Characteristics of energy-level crossings}
In this section we shall discuss the characteristic features
associated with the crossings of the energy levels. Our results
are derived from accurate numerical calculations carried out for
$V_q(r),~q=1-4$ over $\beta=0-100$ in steps of 1.
 We have employed the generalized pseudo-spectral (GPS) Legendre method with mapping, which is a fast algorithm that
 has been tested extensively and shown to yield the eigenvalues with an accuracy of twelve digits after the decimal.
 A more detailed account, with several applications of GPS, can be found in \cite{yao93,cecil01,tong01,roy02,sen06,MAS} and
 the references therein. In the present work, we have also verified the accuracy of these results, in a few selected cases, by using AIM.
 In order to discuss the characteristic features of the level crossings, it is useful to compare the soft-core Coulomb potentials
 with the case of hydrogen atom confined inside an impenetrable spherical cavity of radius $R$.
 The model potential for the spherically confined hydrogen atom ,~SCHA,  \cite{michels37} is given
 by $V_{SCHA}(r)=-\frac{Z}{r} \quad {\rm for}~~r< R;\nonumber ~and~ =~ \infty \quad {\rm for}~~ r\geq R$.~The
 choice of parameters, $R= \infty$ and ${\beta} =0$ in the potentials $V_{SCHA}(r)$ and $V_q(r)$ reduce them to that
 of the free hydrogen-like atom which is characterised by the well known accidental degeneracy of energy levels.
At finite values of the parameters, the accidental degeneracy is
lifted for the two potentials. The relative ordering of $(\nu,l)$,
however, has a different functional dependence on the parameters.
We note here that the level ordering for $V_q(r)$ is similar to
that of muonic atoms where the state with angular momentum $(l+1)$
is more strongly bound than the one with $l$ which is different
from the aufbau principle corresponding to the neutral atoms in
the periodic table. For the two potentials $V_{SCHA}(r)$ and
$V_q(r)$, the higher $l$ states get relatively less destabilized
as $\beta$ or $\frac{1}{R}$ increases (that is to say, the
eigenvalues {\it decrease}). These properties, in conjunction with
the property of monotonicity, give rise to the crossing of a pair
of states $[(\nu,\ell)~(\nu',\ell')]$ with $\nu' > \nu,~l'>l $
which are different for $V_{SCHA}(r)$ and $V_q(r)$. In general,
such crossings produce new degeneracy conditions not present in
the free atom. In the SCHA,  the \emph{simultaneous degeneracy}
condition \cite{pupyshev98,pupyshev02} is obtained such that
\emph{all} $\nu \ge \ell+2$, each $(\nu,\ell)$ SCHA state is
degenerate with $(\nu+1,\ell+2)$ state, when both of them are
confined at $R=(\ell+1)(\ell+2)$. The latter defines the radial
node in the free $(\ell+2,\ell)$ state. For example, at $R= 2\,
a.u.$, \emph{all} pairs of states $(2s,3d), (3s,4d),...$ become
\emph{simultaneously} degenerate. For the soft-core Coulomb potential,
$V_q(r)$, our numerical analysis shows that (a) the crossing takes
place at a certain $\beta$ between a pair of states,
$[(\nu,\ell),(\nu',\ell')]$, defined by $\nu'\geq (\nu+1)$ and
$\ell' \geq (\ell+3)$, (b) $\beta$ varies as $\nu$ changes and it
is not related to the location of radial nodes in the free
hydrogen atom, (d) due to the muonic atom ordering, the higher
than $\ell+3$ states which lie below have already crossed over
before an $[(\nu,\ell),(\nu+1,\ell+3)]$ level crossing occurs. In
Fig.~4 we have displayed the crossings of the $6s$ level with
$7f$,$7g$, and $7i$ levels given by the $V_1(r)$ potential as
$\beta$ is varied.
\begin{figure}[htbp]\centering\includegraphics[width=12cm]{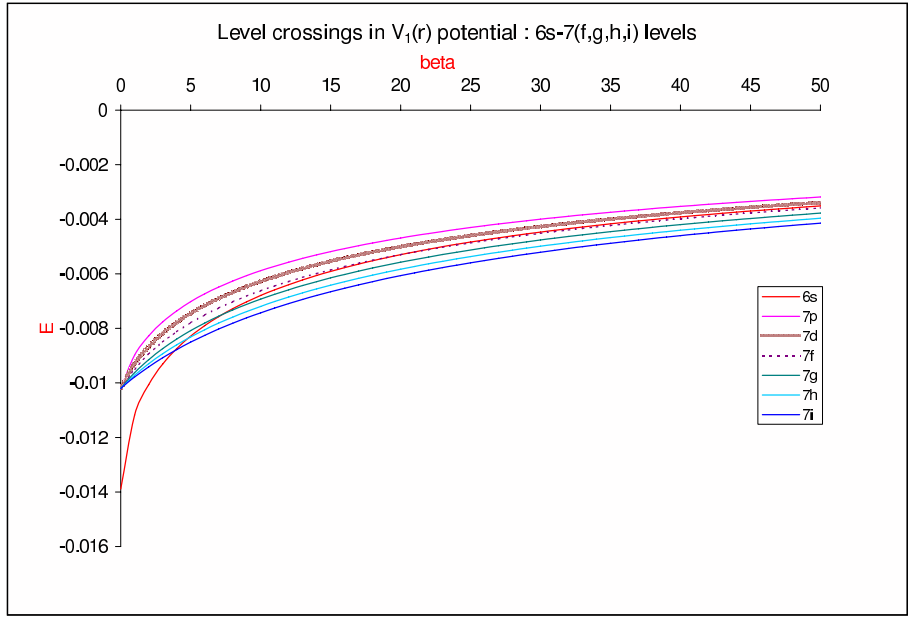}
\caption{Crossings of energy levels in $V_1(r) = -1/(r + \beta)$
for the $6s$ level with $7f$, $7g$, and $7i$ levels as a function
of $\beta$~(in $a.u.$). The energy level ordering over the range
$\beta=20-40$ is changed relative to that corresponding to lower
$\beta$ values.}\label{Fig. 4}\end{figure}

Similar level crossings are obtained in the cases of $V_q(r)$ with
$q=2-6.$ In Fig.~5 shows the level crossings of the $(4p-5g)$,
$(5p-6g)$, $(6p-7g)$ and $(7p,8g)$ pairs derived from $V_2(r)$ as
a function of $\beta$. This potential gives a good approximation
for the laser-dressed hydrogen atom in intense laser field, and
the parameter $\beta$ can be identified with the field strength.
We note in Fig.~5 that the level ordering gets inverted while
passing through the range of the experimentally important region
defined by $\beta=20-40$. In addition to their relevance in
describing the interaction of intense laser fields with atoms
using soft-core Coulomb potential, such theoretical predictions may
have applications in the design of optical devices. Finally, as
noted in $Eq.(4)$, the parameter $q$ is not involved in the energy
scaling. We have also studied numerically the variation of
$E_{\nu\ell}$ as a function of
 $q$ at fixed ${\beta}$ values. We do not observe any
 level crossing in these energy curves.

\begin{figure}[htbp]\centering\includegraphics[width=12cm]{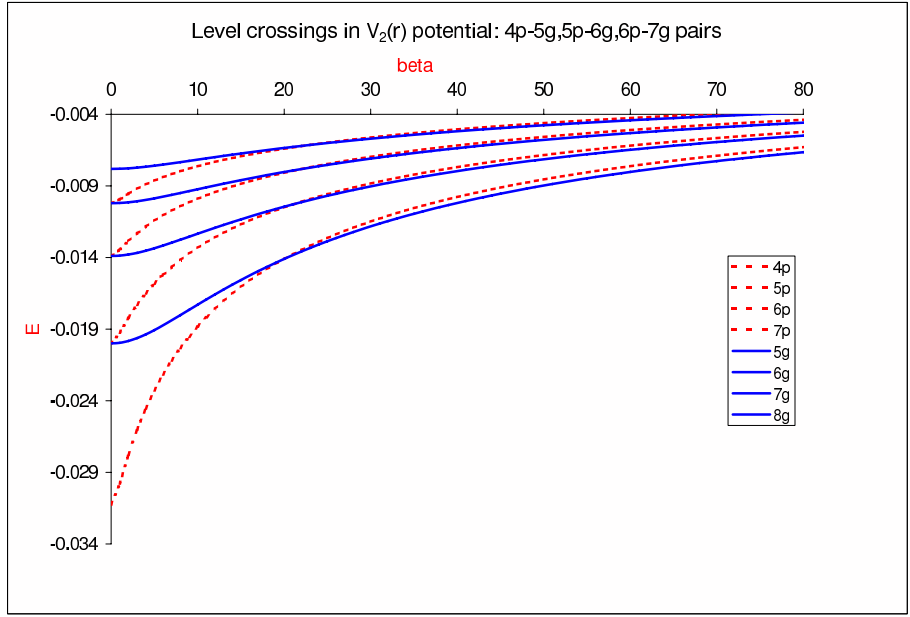}
\caption{Crossings of energy levels in $V_2(r) = -1/(r^2 +
\beta^2)^{1/2}$ for pair of levels $(4p-5g)$, $(5p-6g)$, $(6p-7g)$
as a function of $\beta$~(in $a.u.$). The energy level ordering
over the range $\beta=20-40$ undergoes changes as a consequence of
cross overs.}\label{Fig. 5}\end{figure}

\newpage

\section{Conclusions}
In this work we have obtained several novel analytic results on
the behavior of energies and wave functions for a class of
experimentally useful model soft-core Coulomb potentials represented by
$V_q(\beta,Z,q)$. Since the Hamiltonian $H$ is bounded below, the
energy spectrum can be characterized variationally. This allows us
to conclude that monotonicities of potential in the parameters
$\{Z, \beta, q\}$
 induces corresponding monotonicities in the eigenvalues $E_{\nu\ell}$.
Upper and lower energy bounds have been derived by an application
of the potential envelope method \cite{env1} wherein $V_q(r)$ is
written as a smooth transformation of a monomial potential. Such
bounds are found to obey the same scaling and monotonicity
properties as the exact $E_{\nu\ell}$ and exhibit interesting
geometric property described by a common magnification factor. The
Asymptotic Iteration Method (AIM) is used for the case $V_1$ to
provide exact analytical bound-state solutions for both
eigenvalues and wave functions. We have proved that the electron
density near the nucleus generated from all $V_q$ is \emph{always}
concave. This result, along with the specific condition on $\beta$
corresponding to the exact energy obtained by use of AIM for
$V_1$, and a similar estimate, derived numerically, for $V_2$,
suggests that the ratio $\frac{\eta_l^{\prime\prime} (0)}{\eta_l
(0)}$ contains the information on the identity and location of the
nuclear charge $Z$. Finally, the crossing of energy levels in the
energy spectrum of $V_q,$ have been analyzed using extensive
numerical calculations of $E_{\nu\ell}$ as a function of
${\beta}$. It is conjectured that a pair of states,
$[(\nu,\ell),(\nu',\ell')]$, obey the crossing condition given by
$\nu'\geq (\nu+1)$ and $\ell' \geq (\ell+3)$. This condition is
completely different from that of the hydrogen atom confined
inside an impenetrable spherical cavity, where, at a
characteristic common value of the radius of confinement, all such
pairs of states become \emph{simultaneously} degenerate when
$\ell' = (\ell+2)$. Finally, in the case of the potential $V_2$,
which approximates the laser-dressed hydrogen atom potential, it
is pointed out that the level crossing effects significantly alter
the level ordering over the range of ${\beta} =20-40$, that is
identified with the field strengths of the experimental sources of
intense laser radiation.

\section*{Acknowledgments}
\medskip
\noindent Partial financial support of this work under Grant Nos.
GP3438 and GP249507 from the Natural Sciences and Engineering
Research Council of Canada is gratefully acknowledged by two of us
(respectively RLH and NS). KDS acknowledges the Department of
Science and Technology, New Delhi for the award of J.C.Bose
National Fellowship. KDS and NS are grateful for the hospitality
provided by the Department of Mathematics and Statistics of
Concordia University, where part of this work was carried out.

\end{document}